\documentclass{article}
\usepackage{amsmath,amsfonts}
\usepackage{graphicx}

\usepackage[a4paper]{geometry}

\graphicspath{{figures/}}

\title{Fractional Fourier detection of L\'evy Flights: application to Hamiltonian chaotic trajectories }

\author{\noindent\small Fran\c coise Briolle$^{1,2}$\footnote{Corresponding author, email address:
francoise.briolle@cpt.univ-mrs.fr }, Xavier Leoncini$^2$, Benjamin Ricaud$^3$\\
\small $^1$ CReA, BA 701 Salon de Provence 13300, France. \\
\small$^2$ Centre de Physique Th\'eorique, CNRS-Aix-Marseille Universit\'e,\\
\small Campus de Luminy, Case 907, F-13288 Marseille cedex 9 France.\\
\small$^3$ Institute of Electrical Engineering, EPFL,\\ 
\small CH-1015 Lausanne, Switzerland.}

\begin{document}

\maketitle

\begin{abstract}
A signal processing method designed for the detection of linear (coherent) behaviors among random fluctuations is presented. It is dedicated to the study of data recorded from nonlinear physical systems. More precisely the method is suited for signals having chaotic variations and  sporadically appearing regular linear patterns, possibly impaired by noise. We use time-frequency techniques and the Fractional Fourier transform in order to make it robust and easily implementable. The method is illustrated with an example of application: the analysis of chaotic trajectories of advected passive particles. The signal has a chaotic behavior and encounter L\'evy flights (straight lines). The method is able to detect and quantify these ballistic transport regions, even in noisy situations.
\end{abstract}

\renewcommand{\baselinestretch}{1}
\normalsize

\section{Introduction}
\noindent

The analysis of chaotic signals and the detection of particular patterns inside them are important issues for physics and nonlinear science. The presence of special patterns such as intermittent deterministic behaviors reveal important information on a given physical system.
We propose in this work a signal processing method able to detect regular behaviors occurring in chaotic signals. To demonstrate its efficiency we apply it to signals where L\'evy flights occur: particles display intermittent behavior with 
 almost random motion succeeded with periods of ballistic motion.

This method possesses several key properties required for the study of experimental data. It is robust, not influenced by the nature of the random fluctuations of the signal nor by a reasonable amount of noise which may be present all over the signal (due to experimental measurements). Secondly, it relies on the Fractional Fourier transform. Several numerical implementations of this transform are available among the scientific community which makes the method relatively easy to use for non-expert in signal processing. In addition, this transform can be implemented using fast algorithms.

The robustness of our method relies on an uncertainty principle which is reminiscent of quantum mechanics. It can be shown that one can not measure exactly both frequency and time of a given signal. We use this latter relation to our advantage. Through an elementary transformation we turn random fluctuations of the signal amplitude into random fluctuations of the frequency of a new signal. When these frequencies are rapidly varying, case of random behavior or noise in the signal, the uncertainty principle makes it impossible to have precise information on these variations. In the meantime, coherent behavior is emphasized since it is less fluctuating. As a consequence it eases the detection process and makes it more robust.

This work follows the preliminary results presented in~\cite{RBL} where the use of the uncertainty principle was first stated. We focus here on the signal processing method and integrate it into the more general framework of the Fractional Fourier transform.

In section~\ref{sec:tf} the signal processing method is presented and in
section~\ref{aba:sec1}, we present  an illustration of the method with its application to simulated data from a physical phenomenon: advected particles composed of  random motion and L\'evy flights.

\section{Signals and detection method}\label{sec:tf}

The technique presented here is dedicated to the analysis of signals being made of two ingredients:
\begin{itemize}
\item parts with random fluctuations, e.g. (fractional) Brownian motions, Gaussian or other types of noises.
\item parts with a linear behavior with respect to the variable, often embedded with a reasonable amount of noise ("reasonable" will be made precise in the following).
\end{itemize}
This kind of intermittent signals is typically found in nonlinear physics experiments, for instance in fluid mechanics at the transitions between regular and chaotic/turbulent regimes. An illustration of such signal is shown on Fig.~\ref{fig-phase} (left). Several regions can be distinguished: random fluctuation zones associated to a Brownian motion and some linear regions (of different length and slope) corresponding to a different behavior (L\'evy flights). Note that the linear parts may contain small fluctuations. Our technique is able to detect these linear parts, even embedded in noise, and to measure their length and slope.

\subsection{First step: taking advantage of the uncertainty principle}

In order to follow a rapidly varying signal the measurements must be precise both in the variable value and in the measured quantity depending on it. In some configurations, where the uncertainty principle holds, this is not possible. This principle prevents for example the precise evaluation of the frequency of an oscillating signal when this frequency is  evolving with time (non-stationary signal). It is often a problem in physics but we propose here to use it to our benefit: we need to emphasize the low fluctuation components of our signals among the random variations. For that, we turn our signal into a time-frequency measurement problem.

\begin{figure}[h!]
\includegraphics[width=0.45\linewidth]{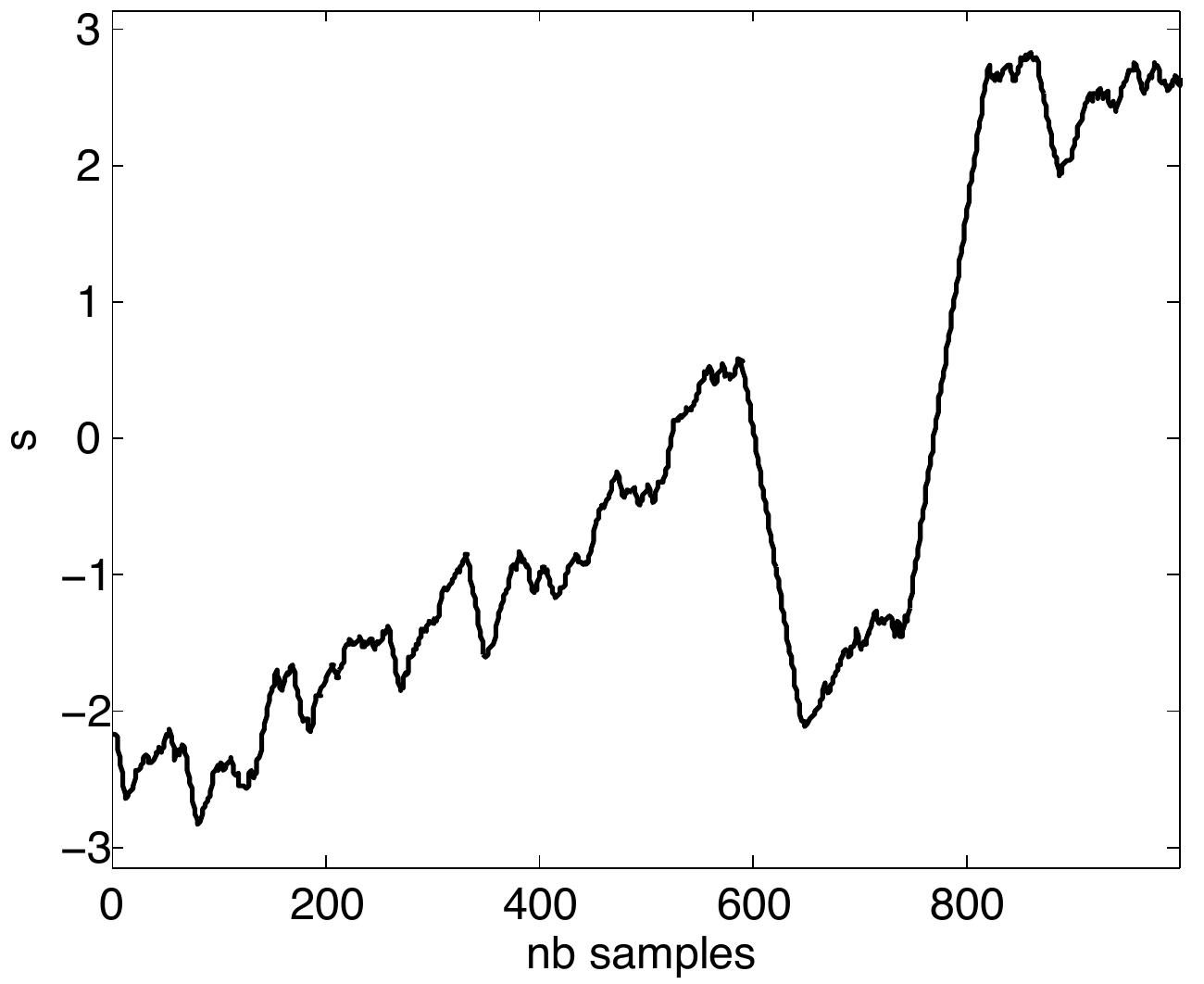} \includegraphics[width=0.5\linewidth]{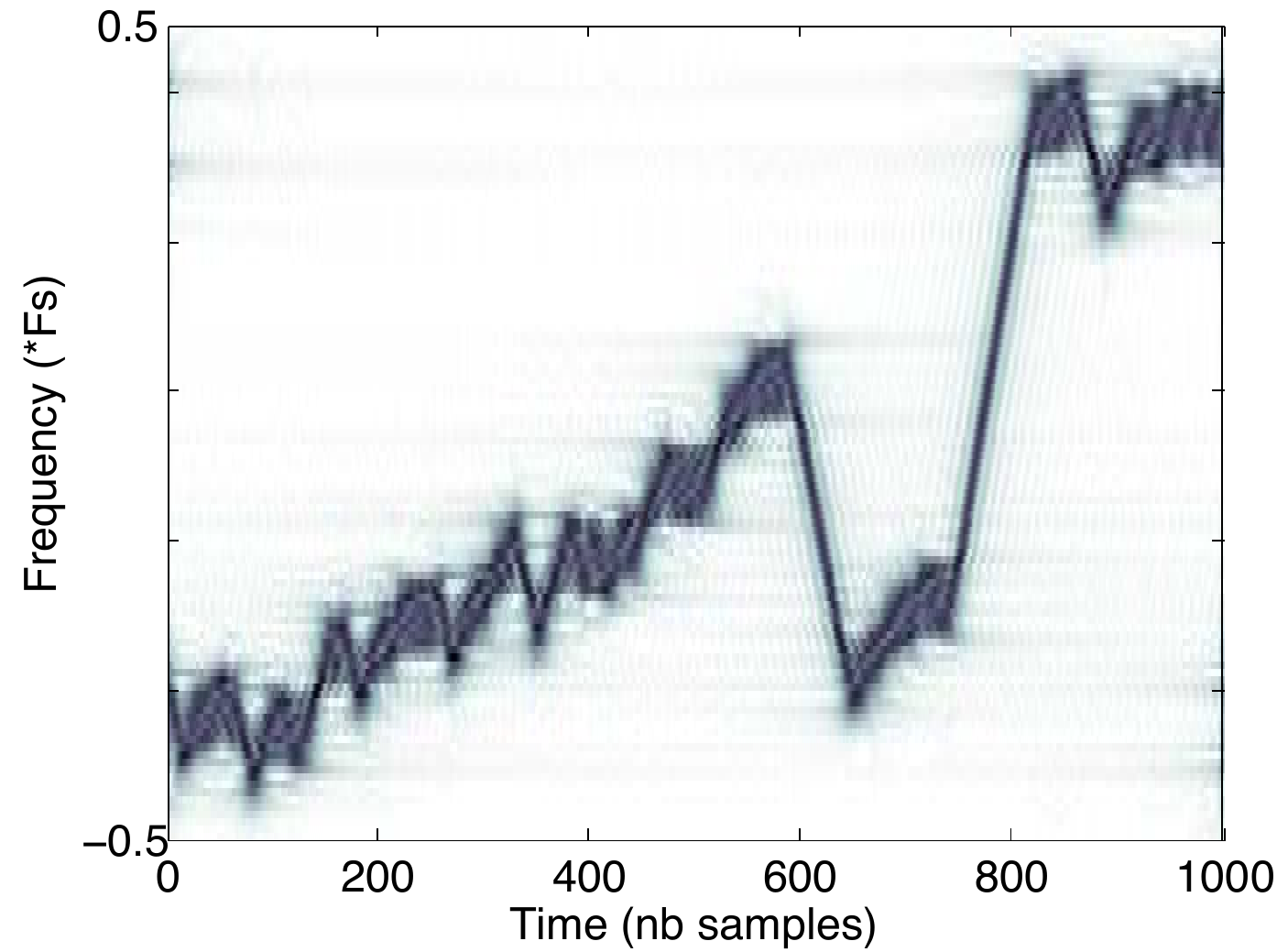}
\caption{Left: tracer trajectory $s$ with fluctuating regions and linear regions (L\'evy flights). Right: spectrogram of $S$ (absolute value of the short-time Fourier transform of $S$). Darker regions are associated to high values of $|{\cal V}_S|$.}
\label{fig-phase}
\end{figure}

 The first step of our analysis is to interpret the signal $s$, a vector of $\mathbb{R}^N$, depending on the variable $t\in [1,2,\cdots,N]$,  as the phase derivative (the fluctuation of the ``frequency component``) of a new signal $S$ depending on time $t$. The oscillating signal $S$ is made of an single non-stationary frequency component in the following way:
\begin{equation}
S(t)=e^{i\varphi(t)},
\label{phase_transform}
\end{equation}
where $\varphi$ is related to the signal $s$ by:
\begin{equation}
\varphi(t)= \sum_{\tau=1}^t s(\tau).
\end{equation}
In order to see the image of $s$ through this transformation, we compute the short-time Fourier transform of $S$:
$${\cal V}_S(t,f)=\sum_{\tau=1}^N S(\tau)e^{2i\pi f\tau} g(\tau-t),
$$
with a Gaussian window $g(\tau)=\exp(-\tau^2)$. The modulus of this representation ${\cal V}_S$ of the function $S$ gives the spectrogram (see e.g.~\cite{Fl} for more details on time-frequency techniques).  As an example,  for $s$ given on Fig.~\ref{fig-phase} (right), $|{\cal V}_S|$ is plotted on Fig.~\ref{fig-phase} (left).
One can guess the signal $s$ from ${\cal V}_S$ but important differences can be seen. First the thin line of the graph of $s$ is now a thick line in the time-frequency plane. Secondly, regions of high fluctuations appear blurred and diffuse. Indeed, the short-time Fourier transform can be seen as a convolution between the modulated signal $S(\tau)e^{2i\pi f\tau} $ and a Gaussian window, which has a blurring action. As a consequence random behavior is blurred even more, spread over a neighborhood zone, whereas linear parts remain relatively sharp. We now turn to the method able to detect and quantify these time-frequency patterns. 

\subsection{Second step: the Fractional Fourier transform (FRTF) for linear frequency component detection}

For the detection of linear behavior in chaotic signals, we need a method able to detect these straight line patterns. In a 2 dimensional image, one would use techniques such as the Hough transform. In our case, we need a similar tool retrieving straight lines which would appear when a time-frequency decomposition is done (such as the short-time Fourier transform, the Gabor transform, or the Wigner-Ville transform). The appropriate tool for this purpose is based on the Fractional Fourier transform. Let us first introduce it. There are two definitions in the discrete setting (where signals are sampled and of finite length)~\cite{FRFT}, which are not equivalent. The first one involves Hermite functions and the second one relies on the discretization of the integral. We choose the second definition as  it is computationally faster than the first one, since it can be done with a fast Fourier transform.
The discrete Fractional Fourier transform ${\cal F}_{\theta}f$ of a signal $f$ of length $N$ is defined as (see~\cite{FRFT,FFRFT}):
\begin{equation}
{\cal F}_{\theta}f(\mu)=\frac{A_{\theta}}{\sqrt{N}} e^{i\pi(\mu^2/N\tan\theta)}\sum_{t=1}^{N} e^{i\left[(\pi/N\tan\theta)t^2-(2\pi\mu/N\sin\theta)t\right]} f(t),
\label{projection32}
\end{equation}
where
\begin{equation}
A_{\theta}=\sqrt{1-\frac{i}{\tan\theta}}=\frac{e^{-i[\pi{\rm sgn}(\theta)/4-\theta/2]}}{\sqrt{|\sin\theta|}},
\end{equation}
The angle $\theta$ parametrizes the transform such that: for  $\theta=0$ we obtain the signal in time (no transform) and for $\theta=\pi/2$ we obtain the standard Fourier transform of $f$. The variable $\mu$ is the time for  $\theta=0$, the frequency for $\theta=\pi/2$ and is a generalized variable for other values of the $\theta$. Since $t\in[1,N]$, $\mu=2\pi n/N$ with $n\in[1,N]$. A connection can be made with the previous study~\cite{RBL} in the following way. Up to a phase factor $\exp(i\mu^2/2\tan\theta)$ and a normalization constant $\sqrt{1-\frac{i}{\tan\theta}}$, the FRFT is the projection of the signal $f$ on a basis of chirp signals:
$$\psi_{\theta,\mu}(t)= e^{i\left[(\pi/N\tan\theta)t^2-(2\pi\mu/N\sin\theta)t\right]}.$$
Notice that applying the FRFT is not strictly equivalent to the calculation done in~\cite{RBL}. The additional phase factor has no importance since only the magnitude of the transform is used to detect the presence of L\'evy flights. However, the normalization factor $A_\theta$ which depends on the angle is important when comparing projections at different angles. Hence the present version (which includes it) is more natural and accurate. Also, one has to change $\mu$ into $(-\mu)$.

As pointed out in~\cite{RBL} and  discussed in~\cite{FFRFT}, numerical instabilities may arise when calculating the Fractional Fourier transform for small values of $\theta$. This correspond to the detection process of L\'evy flights with steep slopes. To cope with this problem, the property ${\cal F}_{\theta-\pi/2}{\cal F}_{\pi/2}={\cal F}_{\theta} $ is used: for $\theta\in[\pi/4,3\pi/4]$ or $\theta\in[-3\pi/4,-\pi/4]$ the FRFT is directly computed and for  $\theta\in[-\pi/4,\pi/4]$ a first Fourier transform is made followed by a FRFT of angle $(\theta-\pi/2)$. This preliminary Fourier transform is equivalent to making a 90 degrees rotation of the time-frequency plane. As a consequence, chirps (L\'evy flights in the frequency-time plane) with large slope get rotated i.e. their slope coefficient gets inverted.

The FRFT is suited for the detection of chirp as $|{\cal F}_{\theta}f|$ will increase whenever a chirp of slope $1/N\tan\theta$  is present. Hence searching for linear patterns in the time-frequency plane is reduced to looking for peaks of the FRFT in the $(\theta,\mu)$ space. Suppose a peak is present at $(\theta_0,\mu_0)$, then the slope $s$ of the linear part can be deduced from $\theta_0$, its shift $d$ from the frequency origin with $\mu_0$ and its length $l$ is proportional to  $|{\cal F}_{\theta_0}f(\mu_0)|$:
\begin{equation}\label{relationchirp}
 s=\frac{1}{N\tan\theta_0},\qquad d=\frac{\mu_0}{N\sin\theta_0},\qquad l=|{\cal F}_{\theta_0}f(\mu_0)| \sqrt{N|\tan\theta_0|+1}.
\end{equation}

The FRFT can be reversed and it is possible to detect a linear part with slope $1/\tan \theta$ inside the signal then erase it in the $(\theta,\mu)$ space and to re-synthesize the signal without this linear part by applying a FRFT of angle $(-\theta)$.

In order to detect the different slopes of the L\'evy flights it is necessary to apply the FRFT for different $\theta$ regularly spaced. The number of selected $\theta$ is fixed by the user depending on how accurate he wants to be and is independent of the length of the signal $N$. The fast implementation of the FRFT is of complexity ${\cal O}(N\log N)$, hence the overall complexity is of the same order.

\subsection{Third step: detection and characterization of L\'evy flights}\label{sec:tf-3}

\begin{figure}[h!]
\begin{centering}
\includegraphics[width=0.5\linewidth]{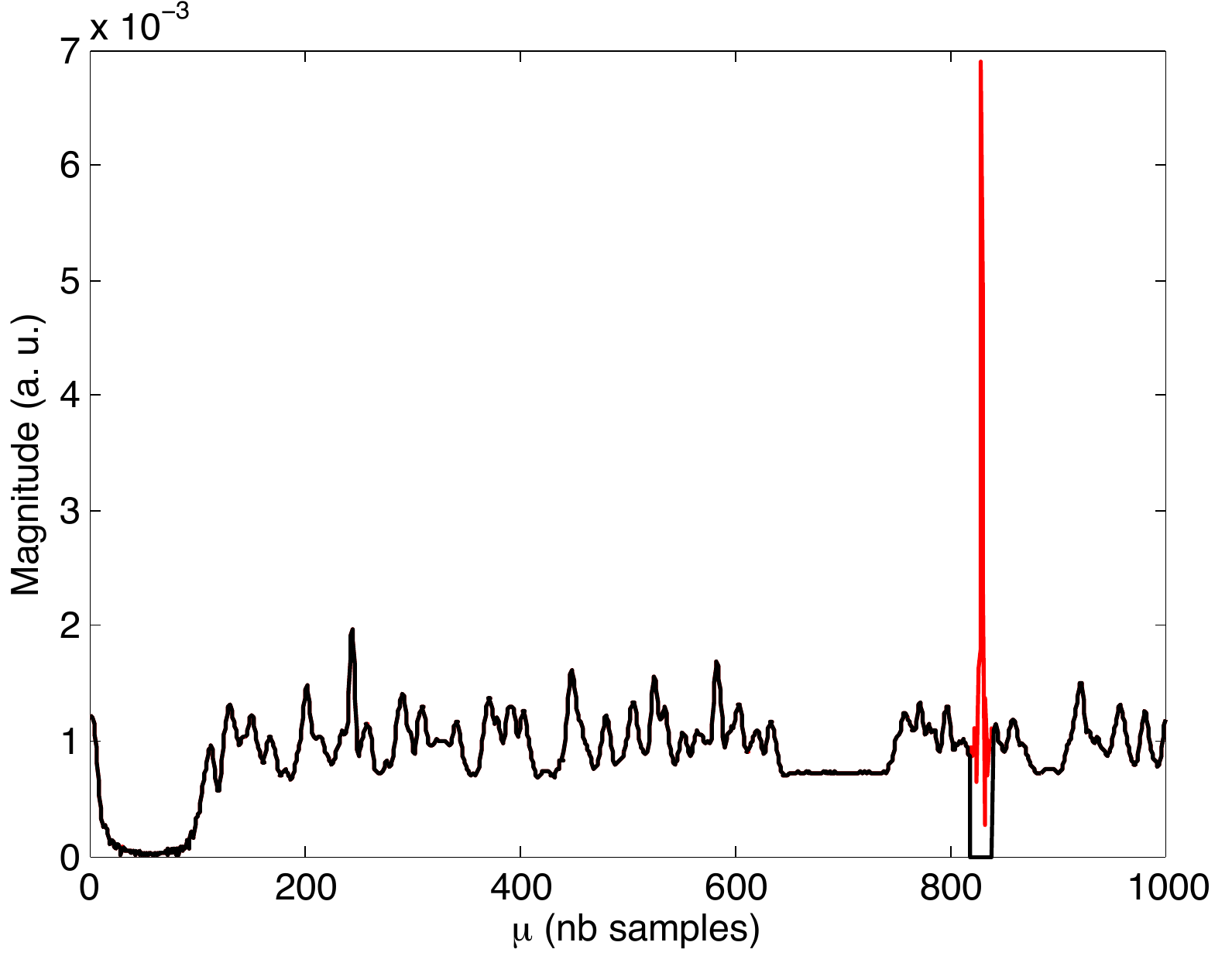}\includegraphics[width=0.53\linewidth]{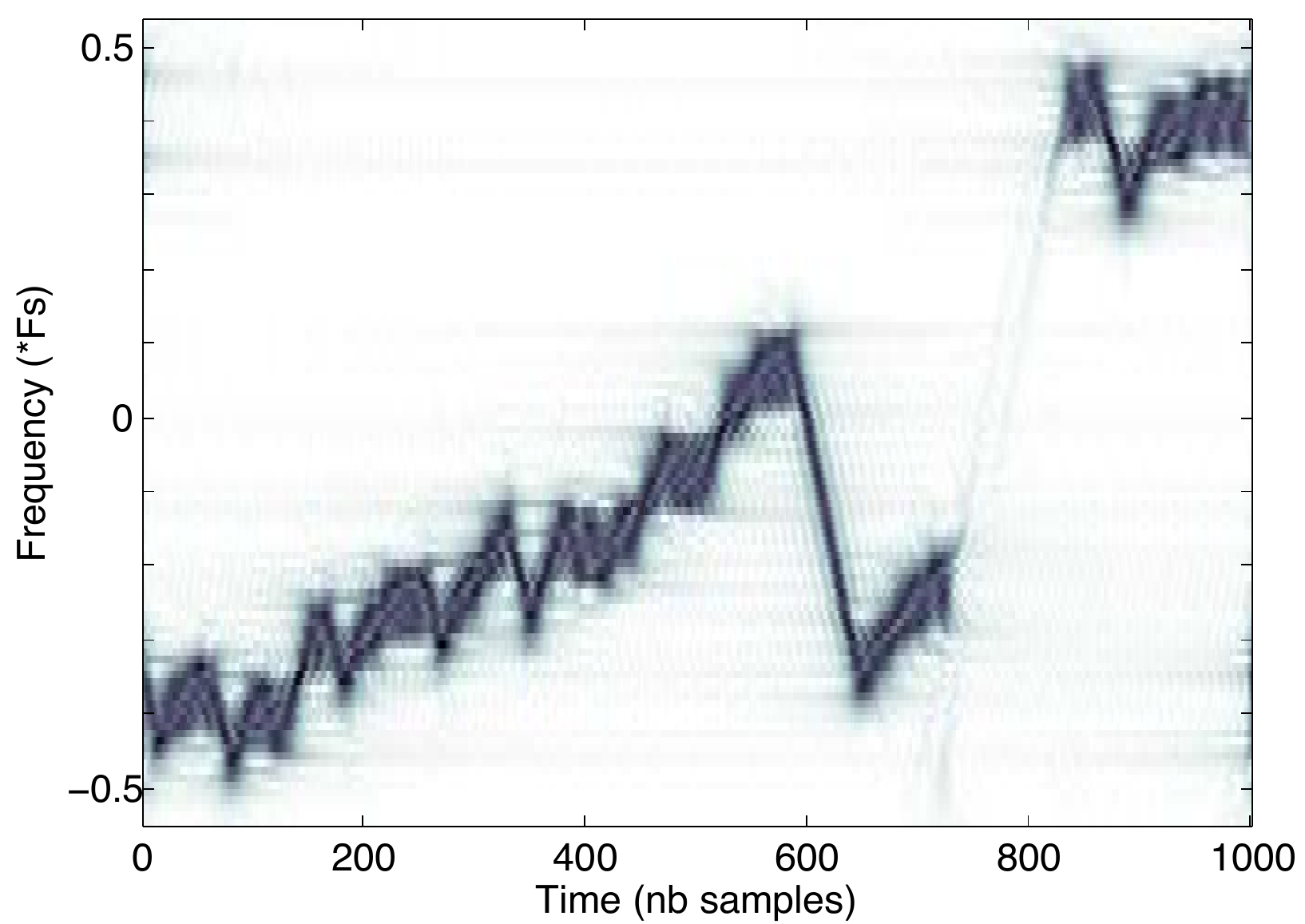}
\par\end{centering}
\caption{Left: for $\theta_m$, signal projections $|C(\theta_m,\mu)|$.  Right : short-time Fourier transform of the signal $S_1$, partial reconstruction of $S$.  The longest L\'evy flight has been removed.}

\label{projection}
\end{figure}

On the signal shown in Fig.~\ref{fig-phase}, one can see several L\'evy flights (left) which have been turned into linear chirps in the frequency-time plane (right). For a specific angle $\theta_m$, the Fractional Fourier transform will produce one sharp peak corresponding to the presence of a chirp. It is illustrated in Fig.~\ref{projection} (left), where $|{\cal F}_{\theta_m}(\mu)|$ is plotted.
For $\mu_m \sim 830$, the sharp peak $|{\cal F}_{\theta_m}(\mu_m)|$ gives evidence that there is a L\'evy flight with a particular slope and length given by the Eq.~(\eqref{relationchirp}].
This search for maxima is the process that detects linear parts in the time-frequency plane.

\vspace{0.25cm}
Since the Fractional Fourier transform is invertible, we can re-synthesize the signal back to the initial representation after setting the values of the transform in red region of Fig.~\ref{projection} (left) to zero. This result is illustrated on Fig.~\ref{projection} (right), which represents the short-time Fourier transform of the newly recreated signal $S_1$. The largest frequency slope of $S$ has been completely removed, the rest remaining untouched. This shows that indeed the peaks in the FRFT correspond to L\'evy flights.

\begin{figure}[h!]
\includegraphics[width=0.5\linewidth]{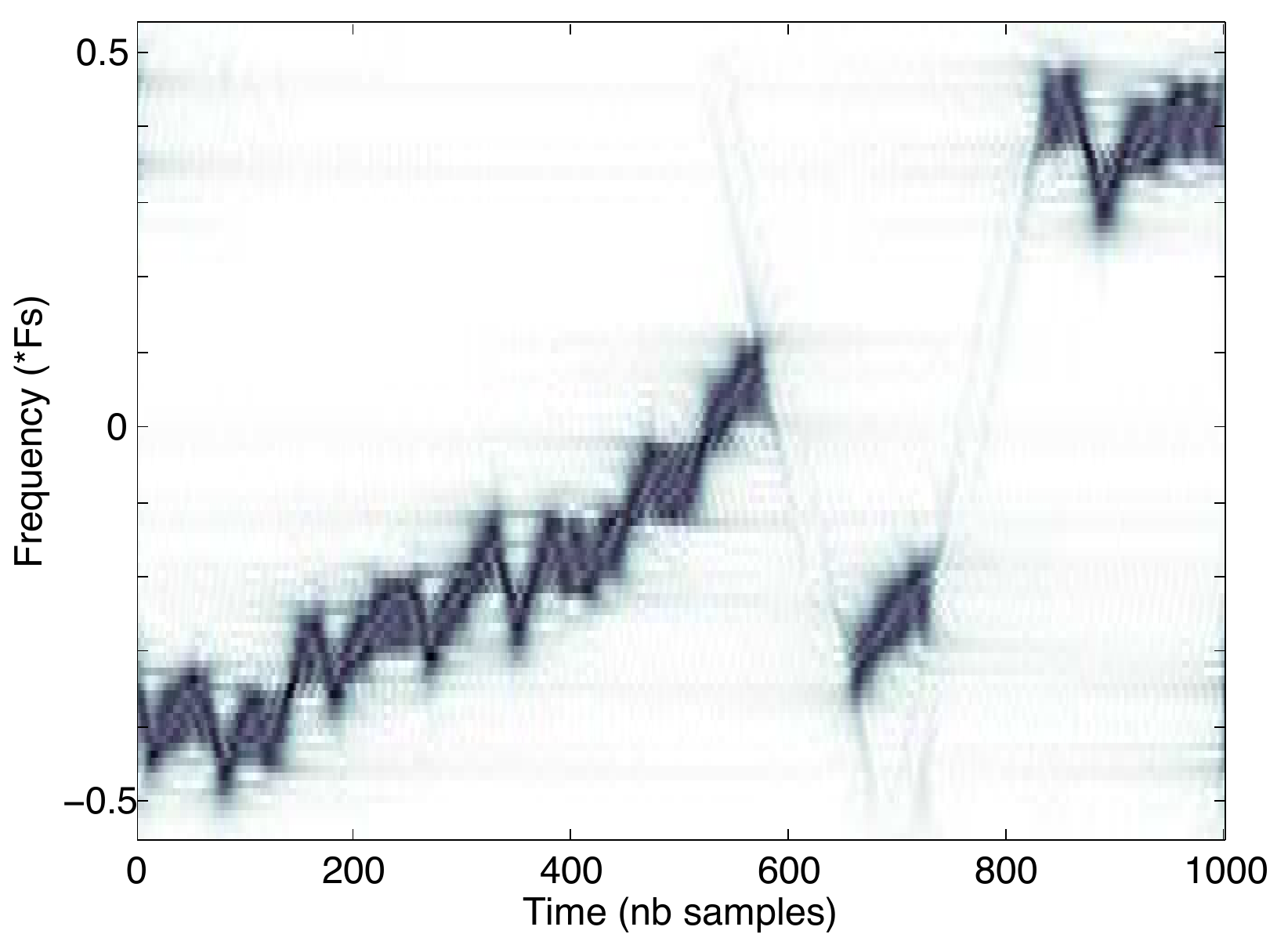}\includegraphics[width=0.5\linewidth]{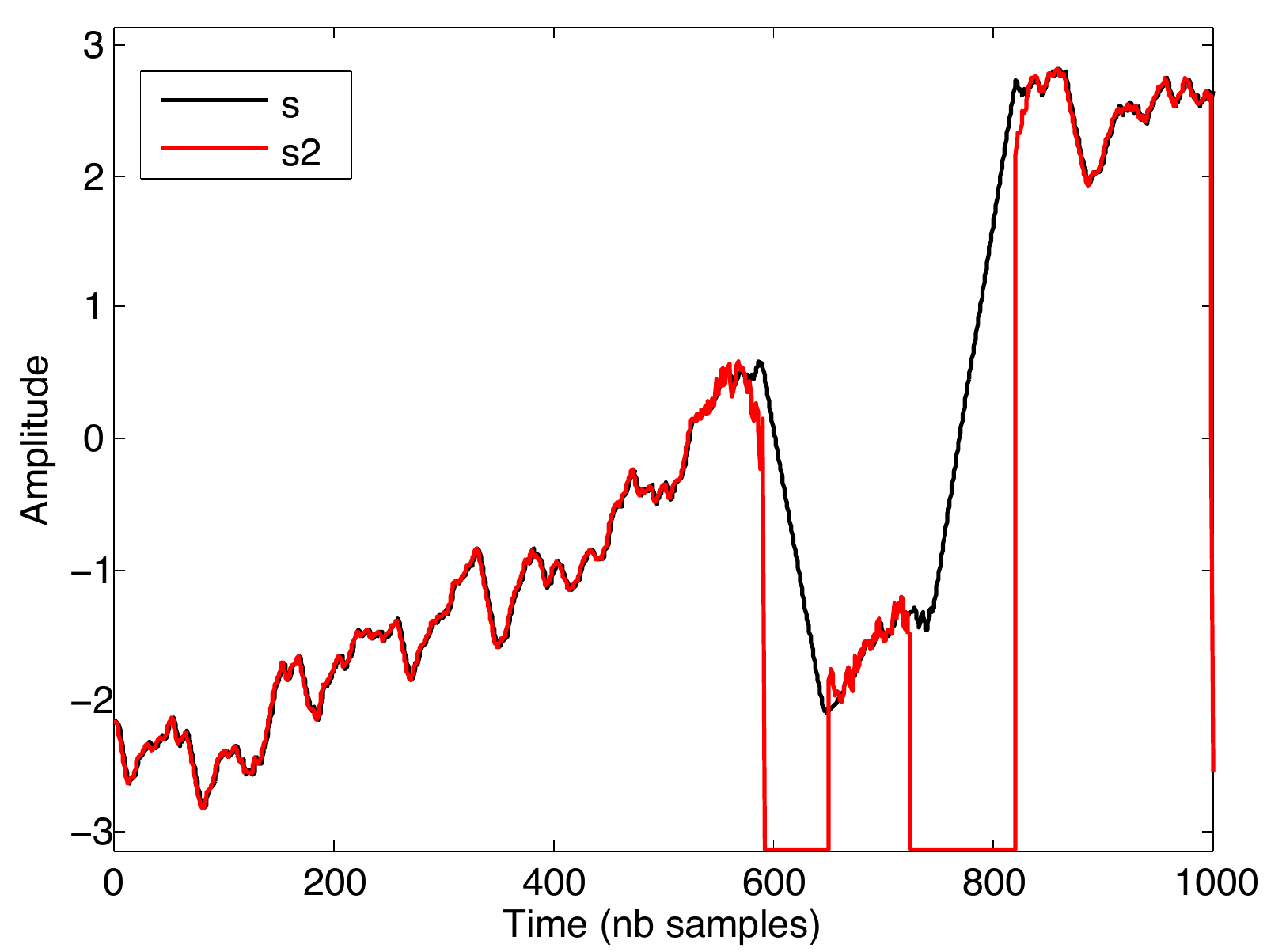}

\caption{Left: short-time Fourier transform of the signal $S_2$ where two  L\'evy flights have been removed. Right : original signal $s$ (black) and partial reconstruction  $s2$, without L\'evy flights (red).}
\label{deux_vols}
\end{figure}

In order to detect all the L\'evy flights, a search of the peaks in the $(\theta,\mu)$ plane has to be done. In \cite{RBL} the suggestion to use a matching pursuit has been proposed. This fits well with their approach i.e a projection of the signal on a set of vectors, the chirp signals. The same result may be obtained with the Fractional Fourier transform as it is unitary. We have to proceed as follows. Each time a peak as been detected, say at $(\theta_m, \mu_m)$, ${\cal F}_{\theta_m}s(\mu_m)$ is set to zero as well as a small neighborhood (user defined) around $\mu_m$. It is illustrated on Fig.~\ref{projection} where the red region is the selected neighborhood to be set to zero. Doing the inverse Fractional Fourier transform will lead to a signal containing all but the chirp component associated to the peak. This process is to be repeated for all peaks. Let $\{M_i=(\theta_i,\mu_i)\}_i $ be the coordinates of the set of peaks in the FRFT domain, and let $\Omega_i$  be a small neighborhood around each $M_i$. Denote by  $\left.{\cal F}_{\theta_i}s\right|_{\mu_i=0}$ the transformed signal where $\mu_i$ and its neighborhood have been set to zero. At iteration $i$, the suppression of the $i$th peak in $s_i$ is given by:
\begin{align}
s_{i+1}={\cal F}_{-\theta_i}(\left. {\cal F}_{\theta_i}s\right|_{\mu_i=0}).
\end{align}
The effect of this process is shown in Fig.~\ref{deux_vols} (left), which represents the short-time Fourier transform of $S_2$. The two longest chirps have been erased from the signal, the rest has been preserved. A reconstruction of the signal $s_2$, which is obtained by direct time derivative of the phase of $S_2$, is plotted  on the right in Fig.~\ref{deux_vols}. Remark that this method do not affect the random part of the signal, only the L\'evy flight parts are removed. We then could use the remaining random part to perform other analysis.

\subsection{Robustness to noise}
In experimental conditions, measurements are always impaired by noise coming from various sources. Hence a method dedicated to the analysis of experimental data must still perform its detection despite a relatively high level of noise. We show here that even with an additional Gaussian noise, our method is still efficient.

\begin{figure}[h!]
\includegraphics[width=0.47\linewidth]{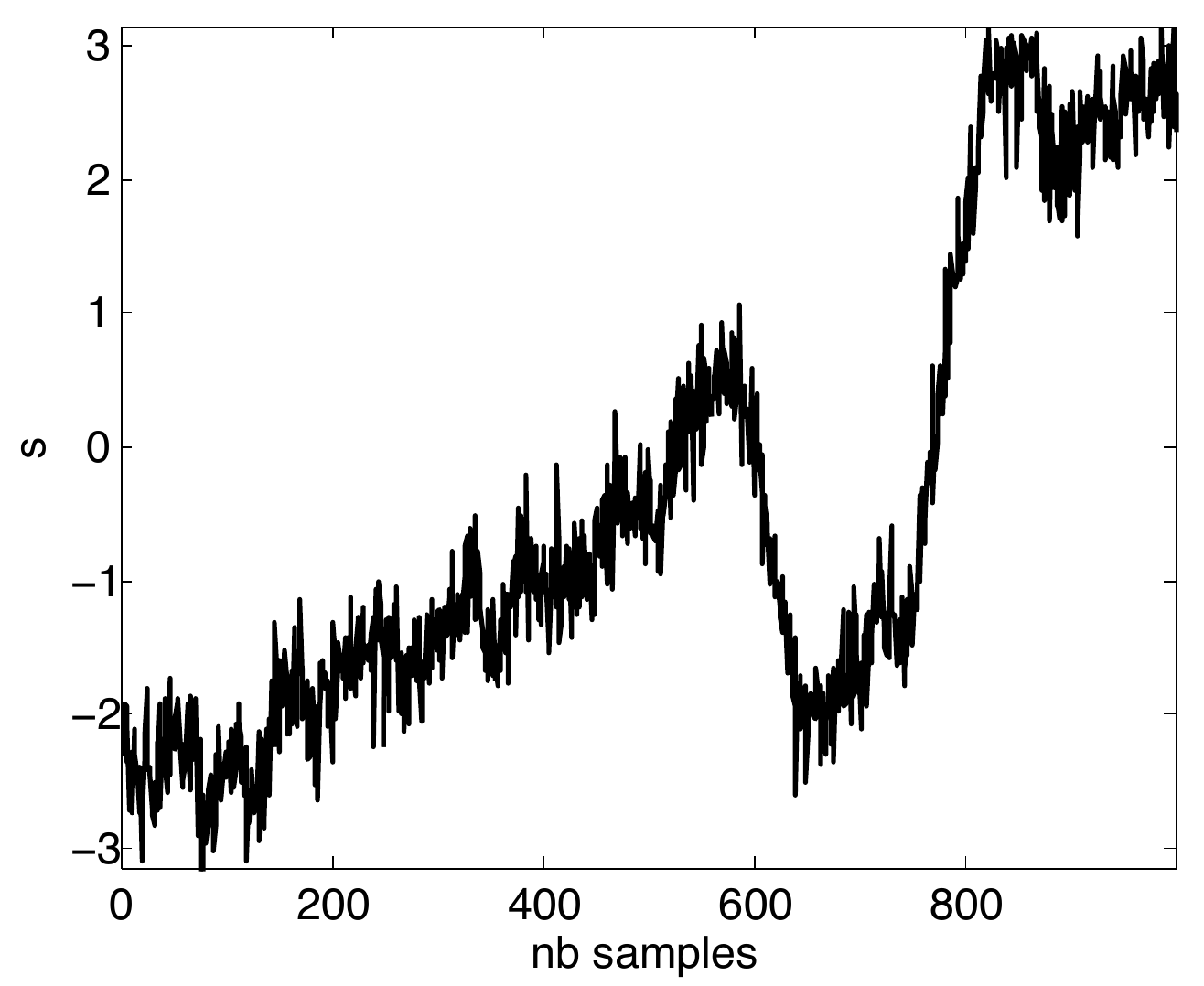}\includegraphics[width=0.52\linewidth]{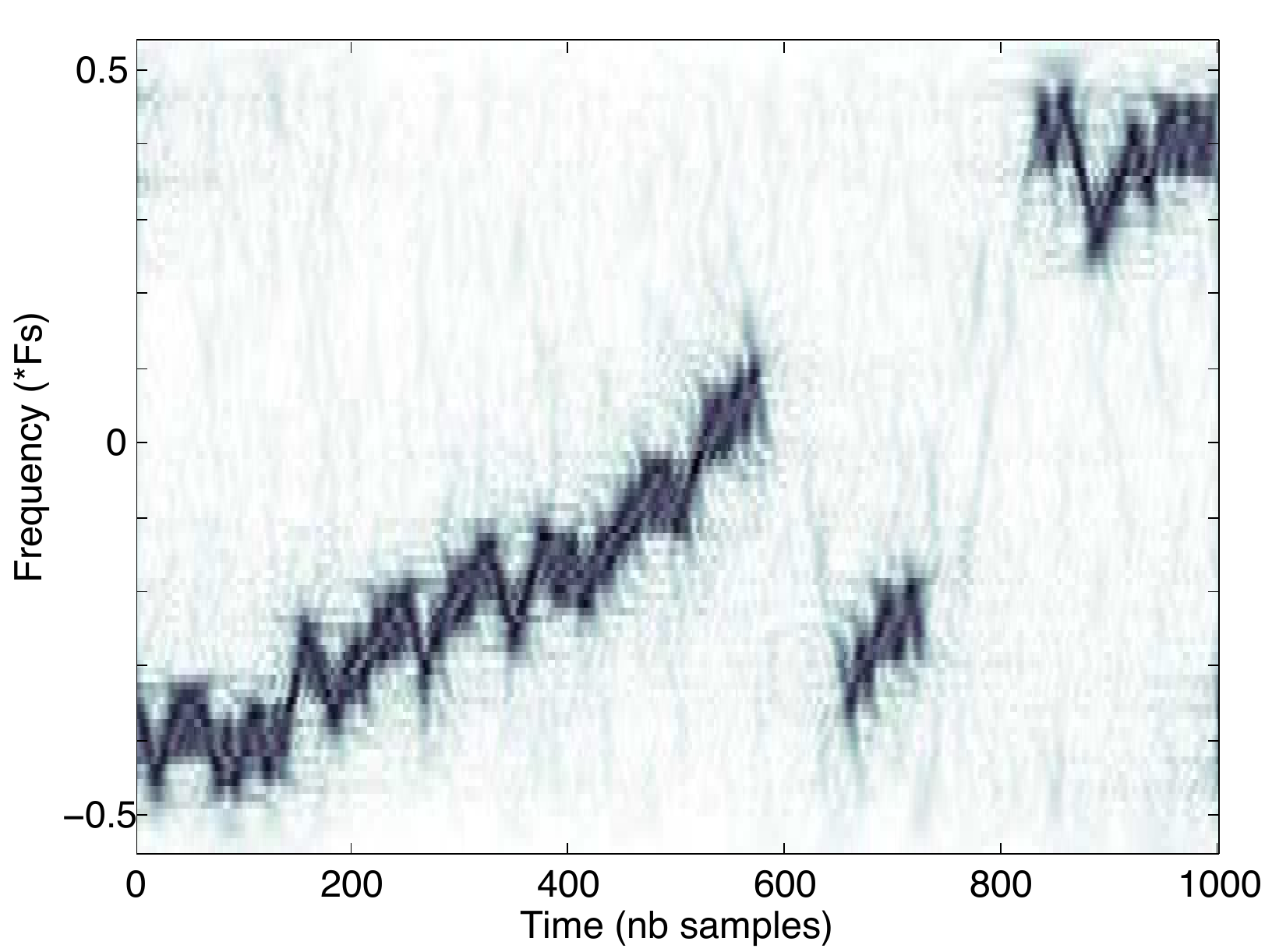}

\caption{Left: tracer trajectory $s$ with Gaussian noise (SNR $\sim$ 17 dB). Right: short-time Fourier transform of the signal $S_2$ where two largest L\'evy flights have been detected and removed. }
\label{deux_vols_bruit}
\end{figure}

The signal to noise ratio (SNR), is defined as:
\begin{align*}
SNR = 10 .\log_{10} \frac{P_s}{P_n}=10 .\log_{10} \frac{\sum s_i^2}{\sum n_i^2},
\end{align*}
where $n$ is the noise. We have added a Gaussian  noise with different levels of amplitude to our signal. The method manages to characterize and extract the two main L\'evy flights for signal to noise ratios down to 17 dB. Fig.~\ref{deux_vols_bruit} shows an example of noisy signal (left) and the resulting signal $S$ after the extraction of the two longest linear behaviors (right).

We shall now apply our method to a specific example as a proof of concept. Namely we shall consider data originating from chaotic advection. Before doing so we shall briefly present the phenomenon and the physical context.

\section{Stickiness and L\'evy flights in chaotic advection}\label{aba:sec1}
\noindent
In this section we briefly discuss the phenomenon of stickiness that occurs in low-dimensional Hamiltonian systems.
Stickiness occurs in the vicinity of some islands of regular motion, inducing
memory effects and L\'evy flights.
In order to be more explicit we shall consider a specific example where this occurs, namely the  phenomenon of chaotic advection of passive tracers. In order to generate a specific flow we shall consider the one generated by three vortices
(see for instance \cite{LKZ01}).

\subsection{Definitions}

For chaotic advection we  consider a flow $\mathbf{v}(\mathbf{r},t)$ of an incompressible fluid ($\nabla\cdot\mathbf{v}=0$).
The notion of a passive particle corresponds to an idealized particle which
presence in the fluid  has no impact
on the flow. This is usually not true, but if the particle is
small enough  this can be a good approximation.
The particle is then just transported by the fluid and its motion is
given by the  equation:
\begin{equation}
\dot{\mathbf{r}}=\mathbf{v}(\mathbf{r},t)\:,\label{eq:gene_advec}\end{equation}
where $\mathbf{r}=(x,y,z)$ corresponds to the passive particle position, and the $\dot{~}$ to a time derivative.

For  a two-dimensional flow,  Eq.~(\ref{eq:gene_advec}) corresponds actually to Hamiltonian equation of motion.
Since the flow is incompressible, we can define a stream function which resumes
to a scalar field, meaning that $\mathbf{v}=\nabla\wedge(\Psi\:\mathbf{z})$,
where $\mathbf{z}$ is the unit vector perpendicular to the considered two
dimensional flow. The equations governing the motion of a passive tracer
Eq.~(\ref{eq:gene_advec})
become
\begin{equation}
\dot{x}=\frac{\partial \Psi}{\partial y},\hspace{1.2cm}\dot{y}=-\frac{\partial \Psi}{\partial x}\:,\label{eq:Hamilton_advec}\end{equation}
where the space coordinates  $(x,y)$ correspond to the
canonical conjugate variables of the Hamiltonian $\Psi$.

When $\Psi$ is time independent, the system is integrable, and particles
follow stream lines. When the stream function  $\Psi$ becomes time-dependent, we
end up with a Hamiltonian system with  $1-\frac{1}{2}$ degrees of freedom. These
systems generically exhibit Hamiltonian chaos. This phenomenon was dubbed
chaotic advection \cite{Aref84,Aref90,Ottino90}. As a consequence chaotic
advection can enhance drastically the mixing properties of the flow, in the
sense that mixing induced by chaotic motion is much more rapid than the one
occurring naturally through molecular diffusion. This is even more important when
the flow is laminar \cite{Ottino89,Zaslav91,Crisanti91}.
When dealing with mixing in micro-fluid experiments and devices  chaotic advection becomes crucial. Indeed since the Reynolds number are usually small,  chaotic mixing becomes, de facto, an  efficient way to mix. There are also numerous  domains of physics,  displaying chaotic advection-like phenomena,  for instance in geophysical flows or magnetized
fusion plasmas \cite{Brown91,Behringer91,Chernikov90,Dupont98,Crisanti92,Carreras03,Annibaldi00,Castillo2004,Leoncini05}.

In order to test the L\'evy detection protocol we established, we  will consider
data originating from passive particles which have been advected by a
two-dimensional flow generated by three point vortices. We shall thus briefly
recall the notion of a point vortex.

\subsection{Point vortex systems}

As mentioned before moving on to advection, let us discuss briefly the flow
generated by point vortices. For this purpose we start with the equation
governing the vorticity of a perfect two-dimensional incompressible flow (the
Euler equation):
\begin{equation}
\frac{\partial\Omega}{\partial t}+\{\Omega,\Psi\}=0\:,\:\Omega=-\nabla^{2}\Psi\:,
\label{vorticity15}
\end{equation}
where $\{\cdot,\cdot\}$ corresponds to
the  Poisson brackets. To get the point vortex dynamics we  consider  a
vorticity field given by a superposition of Dirac functions:
\begin{equation}
\Omega(\mathbf{r},t)=\sum_{i=1}^{N} \Gamma_{i}
\delta\left(\mathbf{r}-\mathbf{r}_{i}(t)\right)\:,
\label{eq:vort_dirac}
\end{equation}
where, $\Gamma_{i}$ designates the strength (vorticity) of a point vortex
located in the two-dimensional plane on the point $\mathbf{r}_{i}(t)$. One
then finds that this singular  distribution becomes an exact solution (in the
weak sense) of the equation~(\ref{vorticity15}) when the
 the $N$ point vortices have a prescribed motion\cite{Machioro94}. To be more
specific the dynamics has to reflect the one originating from  $N$-body
Hamiltonian dynamics.  And when considering no boundary condition, meaning
allowing the flow to live on the infinite plane, the Hamiltonian becomes
\begin{equation}
H=\frac{1}{2\pi}\sum_{i>j}\Gamma_{i}\Gamma_{j}\ln|\mathbf{r}_{i}-\mathbf{r}_{j}
|\:, \label {Hamilton}
\end{equation}
where the the canonically conjugate variables of the
Hamiltonian  are $\Gamma_{i}y_{i}$ and $x_{i}$, and are thus strongly related
to the actual vortex position $\mathbf{r}_{i}(t)$ in the plane.

When actually computing  the equation of motion originating from the
Hamiltonian (\ref{Hamilton}),  (and this how they actually make
sense and were computed) we can notice
that  each vortex is moving according to  the velocity generated by the other
vortices but himself. Having the evolution of the positions of the point
vortices we have as well access to the stream function (the
Hamiltonian governing  passive tracers)
\begin{equation}
\Psi(\mathbf{r},t)=-\frac{1}{2\pi}\sum_{i=1}^{N}\Gamma_{i}\ln|\mathbf{r}-\mathbf
{ r } _ { i }(t)|\:.\label{stream}
\end{equation}
Finally we would like to point out  that the
Hamiltonian of the vortices (\ref{Hamilton}) is invariant by translation and by
rotation.  as a consequence of these symmetries and the associated
conserved quantities, the motion of point vortices becomes chaotic when $N>3$
\cite{Novikov78,Aref80}.

So in order to address chaotic advection, we would like to consider a regular
(laminar) time-dependent flow and we therefore settled for a
flow generated by three vortices. The motion of three point
vortices even though integrable shows a larges variety of behaviors,
quasi-periodic and aperiodic flows are both possible, However in order to
address transport properties,  we are  interested in the asymptotic (large
times) behavior; in order to achieve this we
considered a quasi-periodic motion vortices. Work related to transport for the
case of three identical vortices can be found in
\cite{Kuznetsov98,Kuznetsov2000} and the work with vortices with different signs
from which the  data analyzed in this paper was considered, is
reported in
\cite{LKZ01}.

\subsection{Anomalous transport and Stickiness}

We  have discussed chaotic mixing in a flow generated by three point vortices.
In these  systems, transport can be anomalous. To be more precise, the type of
transport is defined by the value of the characteristic exponent of the
evolution of the second moment.

In summary, transport is said to be anomalous if it is not diffusive in the sense that
$\langle X^{2}-\langle X\rangle^{2}\rangle\sim t^{\mu}$, with $\mu\ne1$:
\begin{enumerate}
\item If $\mu<1$ transport is anomalous and we have sub-diffusion.
\item If $\mu=1$ transport is Gaussian and we have diffusion.
\item If $\mu>1$ transport is anomalous and we have  super-diffusion.
\end{enumerate}

\begin{figure}
\begin{centering}
\includegraphics[width=6cm]{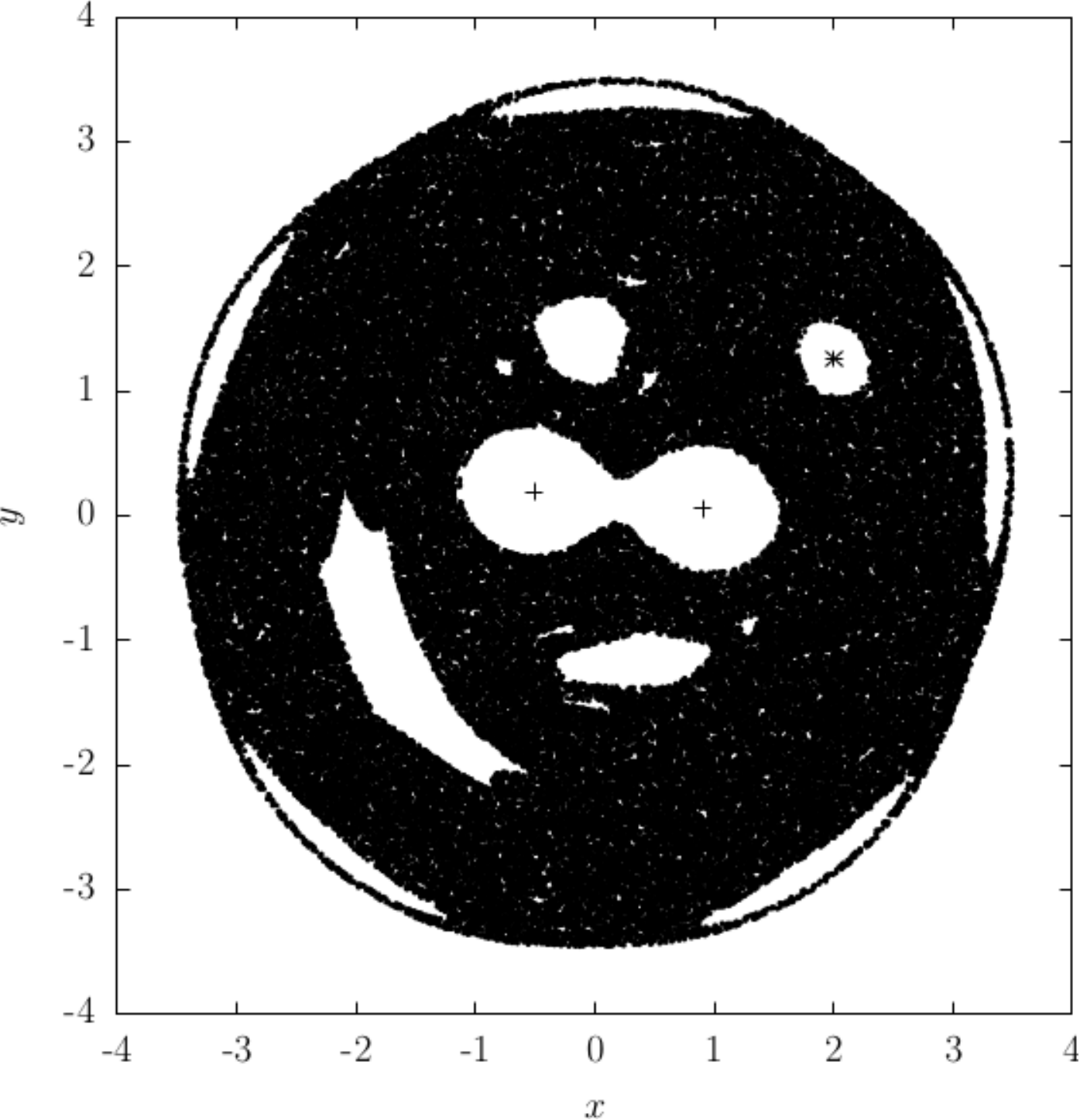}\includegraphics[width=6cm]{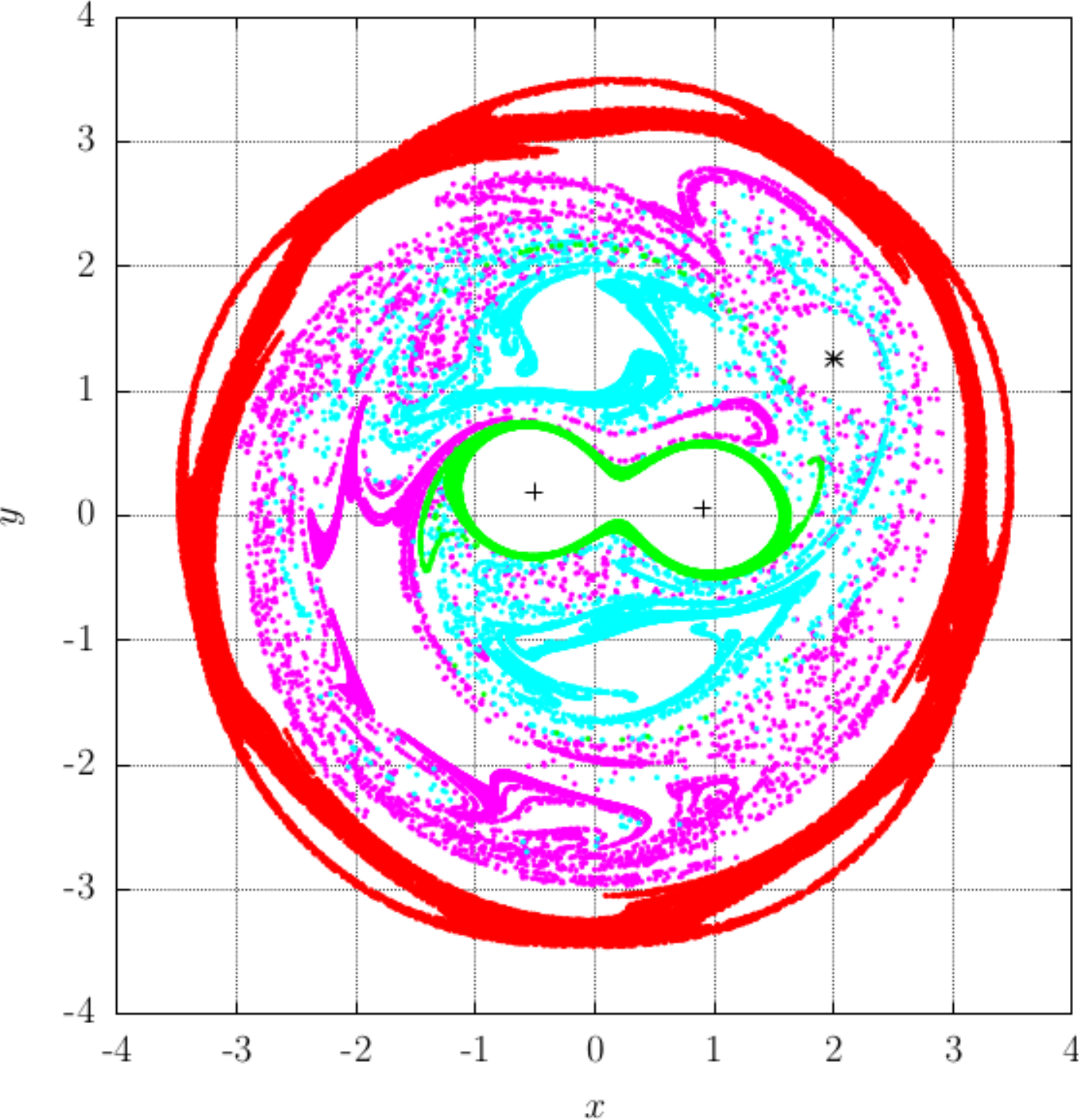}
\par\end{centering}
\caption{Left: Poincar\'e section of the tracer trajectories in the flow generated by three point vortices. Right: localization
of  regions contributing to different types of flights (see \cite{LKZ01} for
details).}\label{poicarresectionk01}
\end{figure}

When considering system of three point vortices, as the one depicted
in Fig.~\ref{poicarresectionk01},
one  notices that the chaotic sea is finite.  Since the chaotic region is  bounded, inferring anomalous properties
is not easy when considering the particle's position, it is easier to work instead with length of trajectories and measure the the dispersion among different trajectories  associated to this quantity
\begin{equation}
s_{i}(t)=\int_{0}^{t}|v_{i}(\tau)|d\tau\:,\label{eq:s(t)}
\end{equation}
where $v_{i}(\tau)$ is the speed of particle $i$ at time $\tau$.
Then to characterize and study transport we compute the moments
\begin{equation}
M_{q}(t)\equiv\langle|s(t)-\langle s(t)\rangle|^{q}\rangle\:,\label{eq:moments}
\end{equation}
where $\langle\dots\rangle$ corresponds to ensemble averaging over
different trajectories.  In order to characterize the transport properties we compute the evolution of the different moments,  from which we extract a characteristic exponent,
 \begin{equation}
M_{q}(t)\sim t^{\mu(q)}\:.\label{mgrowth}
\end{equation}

Transport properties are found to be  super-diffusive and multi-fractal
\cite{LKZ01}, and this is the results of the memory effects engendered
by stickiness: in the vicinity of an island, trajectories can stay for
for arbitrary large times mimicking the regular trajectories nearby, these islands act then  as pseudo-traps. This stickiness   generates a slow decay of correlations (memory effects), which results in  anomalous super-diffusive transport.

To illustrate the phenomenon the Poincar\'e section of passive tracers motion
and the sticking regions are represented in
Fig.~\ref{poicarresectionk01}, (see \cite{LKZ01} for details).
Once a trajectory sticks around  an island,  its length grows almost linearly with time,
with an average speed around the island generically different from the average
speed over the chaotic sea. This implies the presence of  of L\'evy flights in
the data corresponding to trajectories lengths. In
Fig.~\ref{poicarresectionk01}, four sticking regions have been
identified, these regions are naturally expected to give rise to four different
typical average speeds, one therefore expect to identify four different types of
L\'evy flights in the advected data.

\subsection{Multiple signal analysis: blind characterization of L\'evy flights in the advected data}

We now consider blindly data  obtained from the advection of 250 tracers in the point vector flow described in the previous subsection. That is to say, we analyze with our method 250 signals dislaying similar properties as the one presented in section~\ref{sec:tf}. We set up a threshold on the modulus of the projection coefficients (\eqref{projection32}), in order to select only the most relevant L\'evy flights. Similar transport data was 
 was analyzed in \cite{LKZ01}, with traditional tools and found to be anomalous and super diffusive. As mentionned, the starting point of the anomaly was traced back to a multi fractal nature of transport linked to stickiness on four different regular regions. One would thus expect four different type of L\'evy flights in the data (see Fig.~\ref{poicarresectionk01}).

\begin{figure}[h!]
\includegraphics[width=0.5\linewidth]{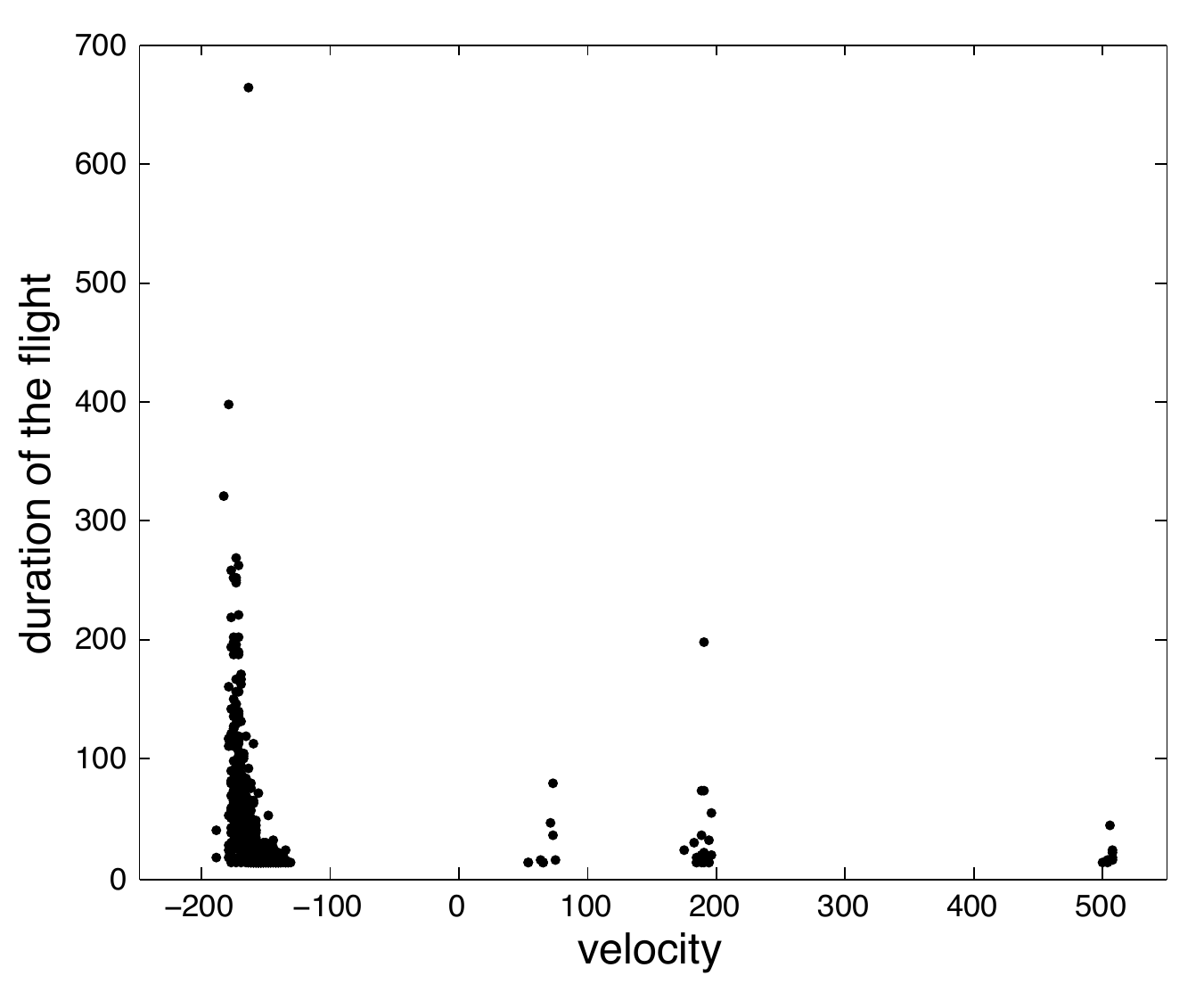}\includegraphics[width=0.51\linewidth]{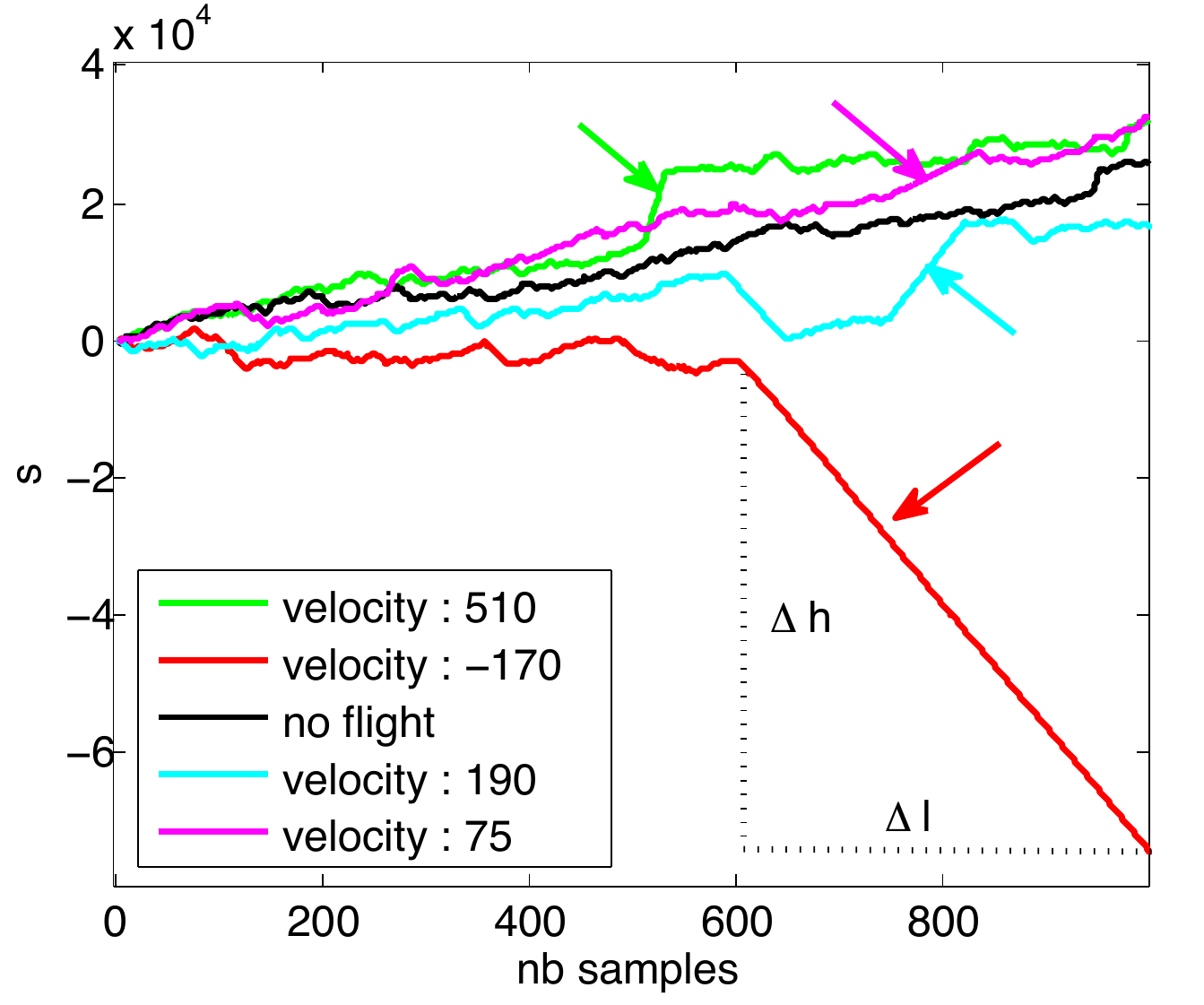}

\caption{Left: duration  of the L\'evy flights as a function of the velocity. Right: the velocity of the main L\'evy flight is plotted for each trajectory.}
\label{velocity}
\end{figure}

In the present case, the method described in part~\ref{sec:tf} has been applied to the data set. Our goal is to detect the multi-fractal nature of the transport resulting from the sticky islands, which would serve as a proof of concept and pave the way to apply the method to numerical and experimental data. The results are presented in Fig.~\ref{velocity}.

For each trajectory, L\'evy flights have been detected and characterized by their length in time, $\Delta l$, and velocity, $\Delta h/\Delta l=s$. The process describe in detail  in part ~\ref{sec:tf-3}, will give, for each flight, its slope (related to the velocity) and length.

The Fig.~\ref{velocity} (left) is an illustration of the duration of the flights as a function of the velocity: four different values have been estimated ($\sim -170, \sim 75, \sim 190$ and $\sim 510$), which means that there are four different types of L\'evy flights, as anticipated. We mention as well that for  some trajectories no L\'evy flights have been detected. A few  few typical trajectories with L\'evy flights have been  plotted on Fig.~\ref{velocity} (right). The color coding corresponds to the one already used in Fig.~\ref{poicarresectionk01}, so that each specific detected flight can be easlily associated to its originating sticky region. 
The agreement with the results found in \cite{LKZ01}, confirm that our method is successful, and is thus ready to be applied  to various numerical and experimental data.

\section{Conclusions}

The first part of the signal processing technique makes use of the uncertainty principle. This has a "dilution effect" on the rapidly varying chaotic parts of the signal while coherent patterns are only slightly affected. This part is critical for the robustness of the detection. Numerical simulations shows that our technique is indeed extremely robust.

The second part of the signal processing technique belongs to the framework of sparsity based analyses. We present a transformation (namely the Fractional Fourier transform) which gives a sparse representation of the data of interest:  L\'evy flights become sharp peaks in the FRFT representation. The key point is that we know the pattern we want to detect and choose the transformation in consequence.

The door is open to further extension and generalization of our method. Suppose one knows \emph{a priori} the patterns to detect which may not be linear but curved or of some other slowly varying shape (slowly varying with respect to the chaotic fluctuations). A different representation than the FRFT should be used based on the shape information. One may use a basis or a set of vectors different from the set of linear chirps. Possible alternatives may be found in e.g.\cite{MMV,BMRV} where what they call "tomograms" are bases of bended chirps and other more general time-frequency forms, associated to one or more parameters (equivalent of $\theta$ in the FRFT case). One may also think of Gabor frames made of chirped windows\cite{BarJ}. Once the representation in which the relevant information is sparse has been found, the peak detection process remains the same.


\end{document}